\documentclass[3p,times,twocolumn]{elsarticle}

\usepackage{ecrc}

\volume{00}

\firstpage{1}

\journalname{Nuclear Physics B Proceedings Supplement}

\runauth{}

\jid{nuphbp}

\jnltitlelogo{Nuclear Physics B Proceedings Supplement}

\usepackage{amssymb}
\usepackage[figuresright]{rotating}

\usepackage{graphicx}

\begin{document}

\begin{frontmatter}



\dochead{}

\title{Sensitivity of the DANSS detector to short range neutrino oscillations}


\author{Mikhail Danilov~$^{a,b,c}$\\
  Representing the DANSS Collaboration (ITEP(Moscow) and JINR(Dubna))}

\address{$^a$ ITEP -- State Scientific Center, Institute for Theoretical and Experimental Physics, Moscow, Russia\\
$^b$ MIPT -- Moscow Institute of Physics and Technology, Moscow Region, Dolgoprudny, Russia\\
$^c$ NRNU MEPhI -- Moscow Engineering Physics Institute, Kashirskoe Shosse 31, Moscow, Russia}

\ead{danilov@itep.ru}

\begin{abstract}
DANSS is a highly segmented 1~$m^3$ plastic scintillator detector. Its 2500 scintillator strips have a Gd loaded reflective cover. Light is collected with 3 wave length shifting fibers per strip and read out with 50 PMTs and 2500 SiPMs. The DANSS will be installed under the industrial 3~GW$_{\rm th}$ reactor of the Kalinin Nuclear Power Plant at distances varying from 9.7 m to 12.2 m from the reactor core.
PMTs and SiPMs collect about 30 photo electrons per MeV distributed approximately equally between two types of the readout. Light collection non-uniformity across and along the strip is about $\pm13$\% from maximum to minimum. The resulting energy resolution is modest, $\sigma /E=15\%$ at 5 MeV. This leads to a smearing of the oscillation pattern comparable with the smearing due to the large size of the reactor core. Nevertheless because of the large counting rate ($\sim$10000/day), small background ($<1\%$) and good control of systematic uncertainties due to frequent changes of positions, the DANSS is quite sensitive to reactor antineutrino oscillations to hypothetical sterile neutrinos with a mass in eV ballpark suggested recently to explain a so-called “reactor anomaly”.
DANSS will have an elaborated calibration system. The high granularity of the detector allows calibration of every strip with about 40 thousand cosmic muons every day. The expected systematic effects do not reduce much the sensitivity region. 
Tests of the detector prototype DANSSino demonstrated that in spite of a small size (4\% of DANSS), it is quite sensitive to reactor antineutrinos, detecting about 70 Inverse Beta Decay events per day with the signal-to-background ratio of about unity. The prototype tests have demonstrated feasibility to reach the design performance of the DANSS detector.
\end{abstract}


\end{frontmatter}

% \linenumbers


\section{The DANSS Detector}
The DANSS collaboration is constructing a relatively compact neutrino spectrometer which does not contain any flammable or other dangerous liquids and may therefore be placed very close to a core of a 3~GW$_{\rm th}$  industrial power reactor at the Kalinin Nuclear Power Plant (KNPP) 350 km NW of Moscow. The size of the reactor core is quite big (3.5 m in height and 3.12 m in diameter). Due to a high $\widetilde{\nu_e}$  flux ($\sim5\times10^{13}\; \bar\nu_e /{\rm cm}^2/{\rm s}$ at a distance of 11 m) DANSS can be used for the reactor monitoring and neutrino oscillation studies. In particular it will be quite sensitive to the reactor $\bar\nu_e$ oscillations to sterile neutrino states with mass splitting of the order of 1~eV$^2$ proposed recently to explain the “reactor anomaly” \cite{Menton}. 

The DANSS detector \cite{DANSS} will consist of highly segmented plastic scintillator with a total volume of 1~m$^3$, surrounded with a composite shield of copper (Cu), lead (Pb) and borated polyethylene (CHB), and vetoed against cosmic muons with a number of external scintillator plates.

\begin{figure*}[ht]
\vspace{-4.0cm}
\hspace*{-2.0cm}
\includegraphics[width=1.17\linewidth]{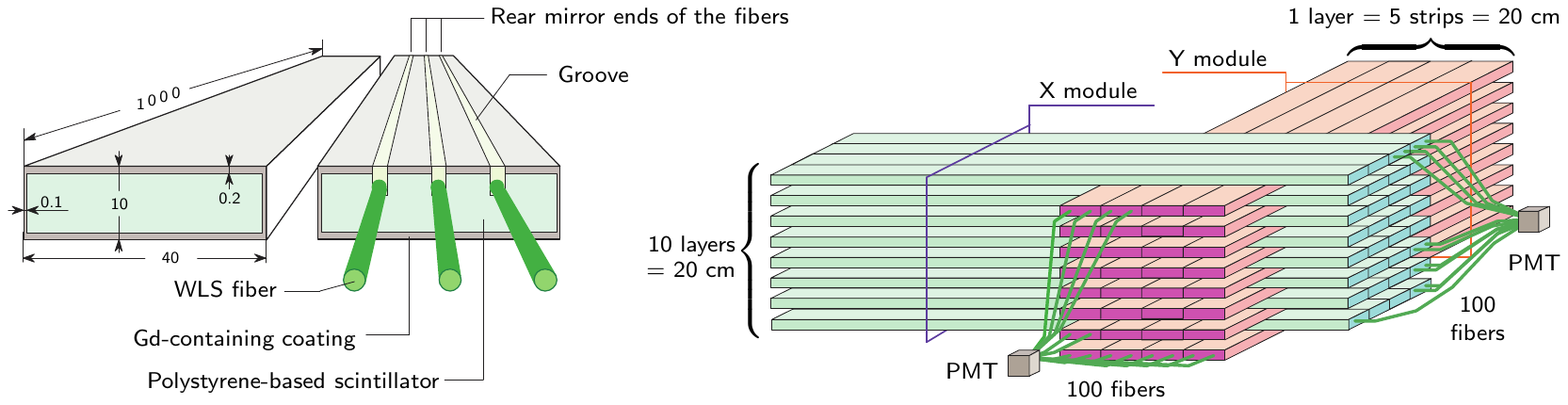}
 \vspace{-19.5cm}
 \caption{\footnotesize The basic element (left) and two of fifty intersecting modules (right) of the DANSS detector.}
 \label{Fig.DANSS_Modules}
\end{figure*}

\begin{figure}[th]
\vspace{-3.7cm}
\centering
\includegraphics[width=1.1\linewidth]{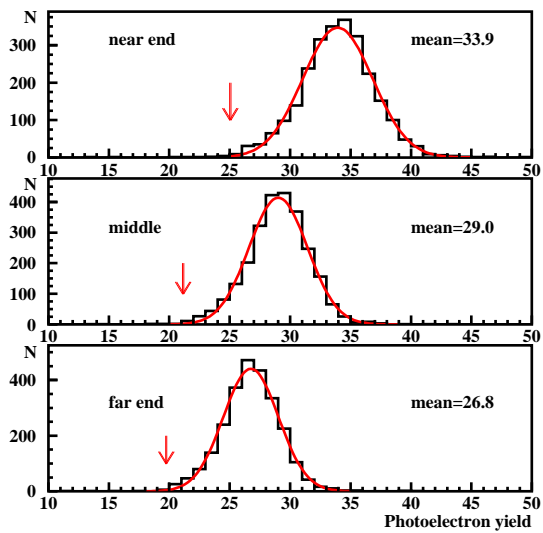}
 \vspace{-0.4cm}
\caption{\footnotesize Photoelectron yield of 2566 scintillator strips at 3 distances from a SiPM:
1.5-26.5cm (top), 38.5-62.5cm (middle), and 73.5-98.5cm (bottom). Arrows show the strip acceptance limits. }
 \label{photo-electrons_vs_L}
\end{figure}

The basic element of DANSS is a polystyrene-based extruded scintillator strip ($1\times4\times100$~cm$^3$) with a thin Gd-containing surface coating which is a light reflector and a ($n,\gamma$)-converter simultaneously (Fig.~\ref{Fig.DANSS_Modules}).
Light collection from the strip is done via three wavelength-shifting (WLS) Kuraray fibers Y-11, $\oslash$~1.2~mm, glued into grooves along the strip. One (blind) end of each fiber is polished and covered with a mirror paint, which decreases a total lengthwise attenuation of a light signal (Fig.~\ref{photo-electrons_vs_L}).

Groups of 50 parallel strips are combined into a module, so that the whole detector (2500 strips) is a structure of 50 intercrossing modules (Fig.~\ref{Fig.DANSS_Modules}). Each module is viewed by a compact photomultiplier tube (PMT) (Hamamatsu R7600U-200) coupled to all 50 strips of the module via 100 WLS fibers, two per strip. 
PMTs are placed inside the shielding but outside the copper layer which serves also as a supporting frame.
In addition, to get more precise energy and space pattern of an event, each strip is equipped with an individual Silicon Photo-Multiplier (SiPM) (MPPC S12825-050C) coupled to the strip via the third WLS fiber. SiPMs are fixed directly at the end of the strip with a plastic light connector.
All signals will be digitized with 12bit, 125MHz FADCs. Only front-end electronics will be placed inside the shielding but outside the Cu layer. All other electronics will be placed outside of the detector shielding. One 6U VME board will serve 64 channels. The maximum trigger rate which can be accepted by the electronics is about 15~kHz. Initially only PMT signals will be used for triggering. Extrapolations of the DANSSino prototype measurements~\cite{EPS,DANSSino} predict acceptable trigger rate of $\sim$1kHz with a simple trigger on the total energy in the detector of larger than 0.25 MeV.

\begin{figure}[th]
\vspace{-2.9cm}
\centering
\includegraphics[width=0.94\linewidth]{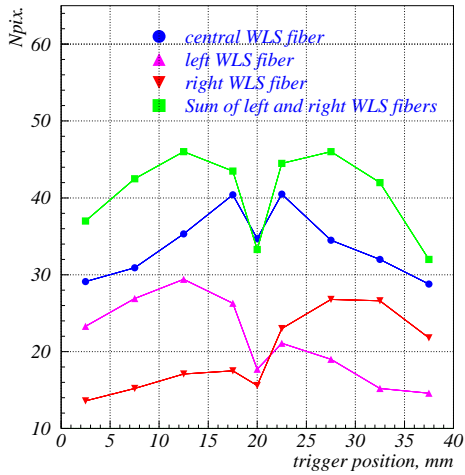}
 \caption{\footnotesize Uniformity of light collection across the strip for the central fiber, two side fibers and their sum, measured with a SiPM. {\it Npix} is a number of fired pixels in the SiPM.}
 \label{Transverse}
\end{figure}

\begin{figure}[th]
\vspace{-3.6cm}
\centering
\includegraphics[width=0.93\linewidth]{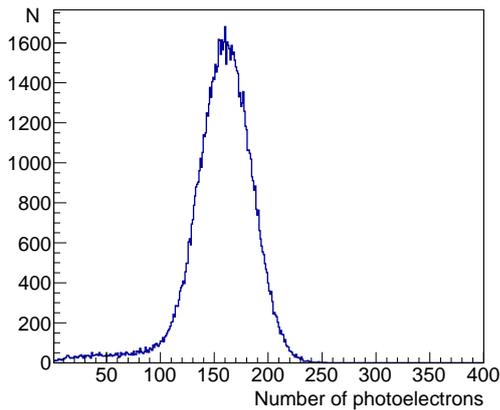}
 \caption{\footnotesize Simulated DANSS response to 5 MeV positrons. }
 \label{resolution5MeV}
\end{figure}

\begin{figure}[th]
\vspace{-5.5cm}
\centering
\includegraphics[width=1.1\linewidth]{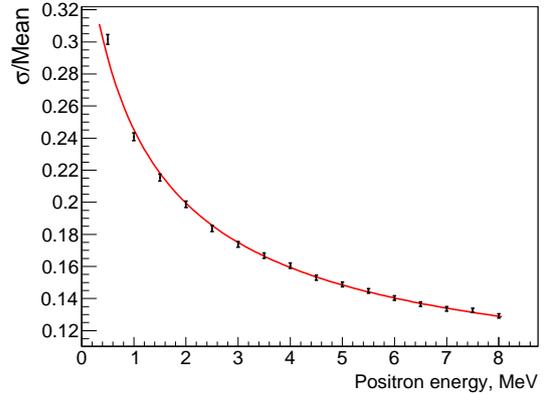}
 \caption{\footnotesize Energy dependence of the DANSS energy resolution. }
 \label{resolution}
\end{figure}

 SiPMs register about 15 photo-electrons (p.e.) per MeV (Fig.~\ref{photo-electrons_vs_L}). This number was obtained using measurements with cosmic rays. Two trigger scintillator counters had the width of 4 cm and the length of 25~cm. The estimated average track length in the strip is 1.1~cm which corresponds to about 2.2~MeV deposited energy. 
The transverse non-uniformity of the scintillator strips is shown in Fig.~\ref{Transverse}. 
It was measured using cosmic rays triggered with 2 thin (0.5 cm width) scintillator counters. The side fibers have a very non-uniform response because of the screening by the central fiber. However the sum of their signals is reasonably uniform.  
PMT sensitivity determined from the calibration source measurements is also about 15 p.e./MeV. So the total number of detected p.e. is 30 p.e./MeV. 
Parameterized strip response non-uniformities have been incorporated into the GEANT4 Monte Carlo (MC) simulation of the detector. Fig.~\ref{resolution5MeV} shows the simulated DANSS response to 5 MeV positron signal.
Energy dependence of the resolution is shown in Fig.~\ref{resolution}.

\begin{figure}[tbhp]
\centering
\vspace{-1.1cm}
\hspace*{-2.0cm}\includegraphics[width=1.6\linewidth]{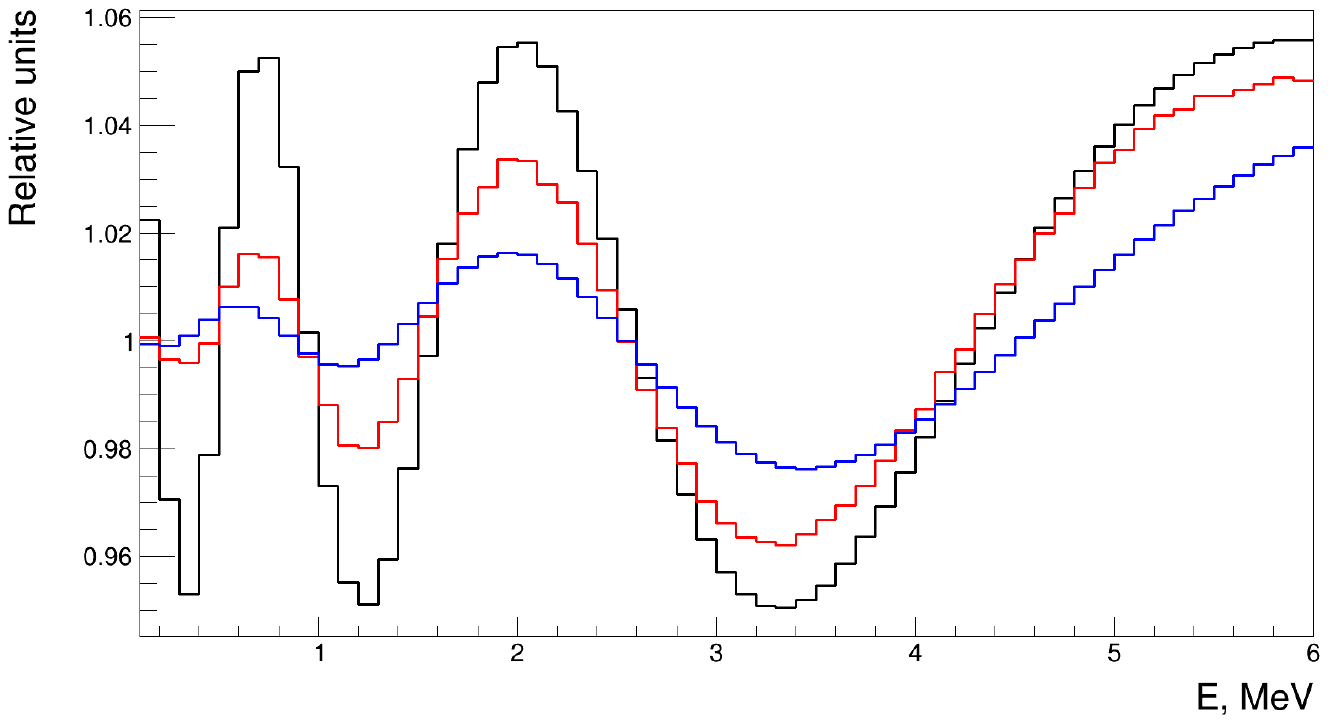}
\vspace{-12.5cm}
\caption{Ratio of energy spectra with neutrino oscilations ($\Delta M^2=2eV^2$, $sin^2(2\theta_{14})=0.1$) and without oscilations, for the distance of 9.7~m from the reactor core in ideal case (black), with length uncertainty (red), and with the length and energy uncertainty (blue). }
\label{Fig.Smearing}
\end{figure}

\begin{figure}[tbhp]
\includegraphics[width=0.95\linewidth]{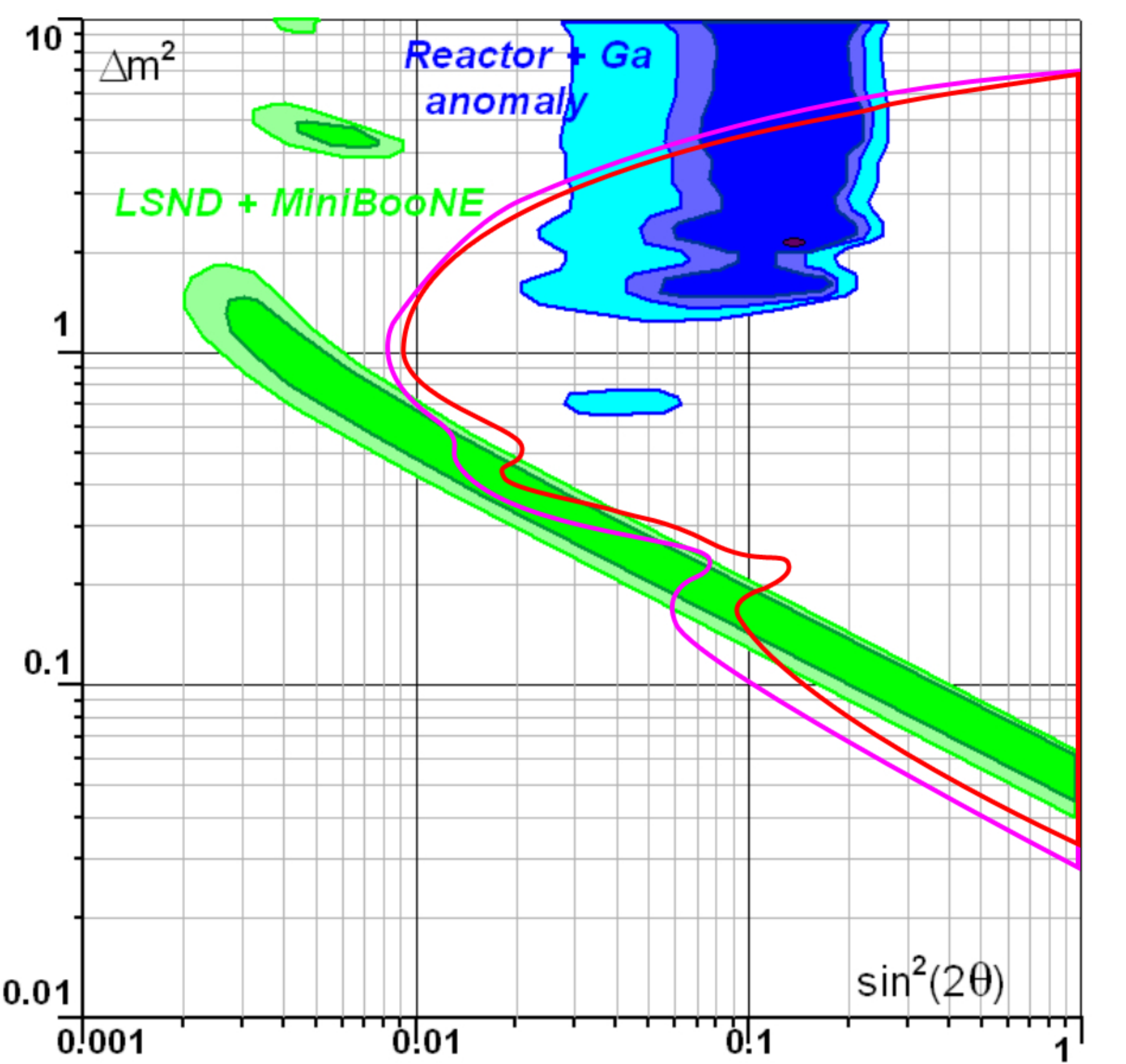}\centering
%\vspace{-6.2cm}
\caption{DANSS 95\% CL sensitivity contours for one year of running at 9.7~m and 12.2~m distances with equally shared statistics, obtained using only shape information of the experimental energy spectra (magenta). Red line shows the sensitivity with systematic uncertainties included. }
\label{Fig.Sensitivity}
\end{figure}

\section{The DANSSino tests}

In order to verify the DANSS design, to measure the real background conditions and shielding efficiency, a simplified pilot version of the detector (DANSSino) was constructed and tested \cite{EPS, DANSSino}. It had similar structure but contained only 100 scintillator strips (4\% of the DANSS).  
The DANSSino detector was installed under the WWER-1000 reactor at the KNPP at a distance of 11 m from the core center. In spite of the small size and non-perfect shielding the DANSSino detected about 70 Inverse Beta-Decay (IBD) events per day with the signal to background ratio of about 1. These tests confirmed the estimates of the DANSS performance.  

\section{DANSS sensitivity to short range neutrino oscillations}
%\section{DANSS sensitivity to neutrino oscillations}

The agreement between MC expectations and actual DANSSino data provides confidence in the MC simulations of the DANSS detector and reliability of the estimated sensitivity to neutrino oscillations. DANSS efficiency for IBD events is estimated to be about 70\%. The antineutrino counting rate will be as high as 10 thousand events per day. At the same time the background level is expected to be about 1\% only. These features together with a possibility to change frequently the distance from the reactor core make the DANSS very sensitive to neutrino oscillations. 

The DANSS modest energy resolution ($\sim$24\% at 1~MeV) leads to the smearing of the oscillation pattern comparable to the smearing due to the large size of the reactor core (Fig.~\ref{Fig.Smearing}). Nevertheless the high neutrino counting rate and a small background lead to a high sensitivity of the DANSS to short range neutrino oscillations proposed to explain the “reactor anomaly”. Fig.~\ref{Fig.Sensitivity} shows the expected 95\% CL exclusion plot obtained assuming one year of running with equal number of events recorded at distance of 9.7 and 12.2 meters form the reactor core (distances are from the core center to the detector center). Sensitivity was estimated using the experimental energy spectra shapes 
information only. The sensitivity can be improved by adding the rate information. The sensitivity can be further improved, in particular in the large mass splitting region, by adding information on the predicted reactor $\bar\nu_e$ spectrum \cite{EPS}. However at the moment it is hard to estimate the uncertainties in the $\bar\nu_e$ spectrum and therefore we do not use this information.
The systematic uncertainties are expected to be very small because of a small background and frequent changes of the detector distance to the reactor core which takes only 5 minutes. In addition frequent calibrations of the detector are foreseen using different radioactive sources that can be inserted in the detector using special tubes. 
Moreover each strip will be continuously calibrated with cosmic muons which span the most interesting energy range from 2 to 5 MeV. About 40 thousand muons per day will be used for calibration of each strip. The direction and position of every muon can be well determined because of the high granularity of the detector. Muons will deposit 20-50 MeV in 10 layers of strips readout by one PMT. Such calibration is also useful although the covered energy range is an order of magnitude higher than positron energies from IBD.
The influence of systematic uncertainties on the sensitivity of DANSS to neutrino oscillations was estimated by allowing at one distance from the reactor core a change of the energy scale by up to 1\% and by adding up to 1\% of background distributed as E$^{-2}$. The reduction of the sensitivity is not large (see Fig.~\ref{Fig.Sensitivity}). DANSS is sensitive to the most interesting parameter region indicated by the reactor and Ga anomalies as shown in Fig.~\ref{Fig.Sensitivity}.

\section{Conclusions}

The DANSS detector will start data taking under the core of the WWER-1000 reactor of the KNNP in the beginning of 2015. Due to the high counting rate ($\sim$10000/day), small background ($<1\%$) and a good control of systematic uncertainties due to frequent changes of positions, the DANSS will be sensitive to the most interesting part of the parameter space for reactor antineutrino oscillations to hypothetical sterile neutrinos suggested to explain the reactor anomaly. 

\section{Acknowledgements}
We are grateful to the staff of the Kalinin Nuclear Power Plant and especially to the KNPP Directorate for providing a possibility of performing research measurements extremely close to the reactor core and to the KNPP Radiation Safety Department for the constant technical assistance.
The work was supported in part by the corporation Rosatom (state contract H.4x.44.90.13.1119) and by the Russian Ministry of Education and Science (grant 4465.2014.2).

\end{document}